\documentclass{PoS}
\usepackage{amssymb}
\usepackage{booktabs}
\usepackage{multirow}
\usepackage{rotating}
\pdfoutput=1

\title{Flavor asymmetry of the polarized nucleon sea}

\ShortTitle{Flavor asymmetry of the polarized nucleon sea}

\author{\speaker{Emanuele Roberto Nocera}\\
        Universit\`{a}  degli Studi di Milano \& INFN Milano - 
        Via Celoria 16 Milano, Italy\\
        E-mail: \email{emanuele.nocera@unimi.it}}

\abstract{We discuss the flavor asymmetry of polarized light antiquarks in the nucleon,
$\Delta\bar{u} - \Delta\bar{d}$. We present a determination of this quantity based on 
two global QCD analyses of experimental data, \texttt{DSSV08} and \texttt{NNPDFpol1.1},
in which sea-quark parton distribution functions are determined respectively either
from semi-inclusive deep-inelastic scattering or $W$-boson production data.
The latter have become available only very recently, and their effect on the 
polarized flavor asymmetry is presented here for the first time. 
We find that the flavor asymmetry of polarized antiquarks in the nucleon is definitely positive, 
and has almost the same absolute size as its unpolarized counterpart.
We compare this result with various theoretical models of the nucleon structure
in order to test their validity. We show that some of them are clearly disfavored.}

\FullConference{XXII. International Workshop on Deep-Inelastic Scattering and Related Subjects,\\
		28 April - 2 May 2014\\
		Warsaw, Poland}

\begin{document}

The flavor dependence of quark and antiquark 
Parton Distribution Functions (PDFs) in the nucleon is presently 
recognized to play a leading role in unraveling the non-perturbative
chiral dynamics of Quantum Chromodynamics (QCD)~\cite{Garvey:2001yq}.
A sizable flavor asymmetry $\bar{u} - \bar{d}> 0$ is observed
between light helicity-averaged antiquark distributions from charged-current 
Deep-Inelastic Scattering (DIS) and Drell-Yan (DY) 
measurements~\cite{Peng:2014hta}. This leads to the question whether 
their helicity-dependent counterparts, $\Delta\bar{u}$ and $\Delta\bar{d}$, 
would be asymmetric as well. The knowledge of such 
a behavior is of particular relevance in the polarized case, 
since it could provide additional insight into 
the decomposition of the proton's spin into quark and antiquark contributions.

Polarized light-sea distributions can be accessed experimentally in
different hard-scattering processes: polarized Semi-Inclusive DIS
(SIDIS), polarized DY~\cite{Bunce:2000uv}, single-spin asymmetry in $W$-boson
production in polarized proton-proton (\textit{pp}) collisions~\cite{Bourrely:1993dd},
and charged-current DIS using both neutrino beam, at a neutrino
factory~\cite{Mangano:2001mj}, or electron beam, at an Electron-Ion 
Collider~\cite{Boer:2011fh}. Among these processes, only SIDIS and $W$-boson
production have been actively pursued so far, and some of the corresponding 
experimental data have been included in global QCD analyses of polarized PDFs.
The \texttt{DSSV08}~\cite{deFlorian:2009vb} and 
\texttt{LSS10}~\cite{Leader:2010rb} determinations of PDFs 
include SIDIS data, while the \texttt{NNPDFpol1.1}~\cite{Nocera:2014gqa} 
determination of PDFs includes $W$-boson production data.

Various theoretical models have also been developed for predicting the 
unpolarized and polarized flavor structure of the nucleon sea  
(for a review, see \textit{e.g.}~\cite{Chang:2014jba}). 
Computations based on different models lead to similar results for the 
unpolarized distributions $\bar{u}$ and $\bar{d}$, while they often differ 
significantly for the polarized distributions $\Delta\bar{u}$ 
and $\Delta\bar{d}$. Therefore, a comparison between 
the latter predictions and the corresponding
PDFs determined from a fit to experimental data allows 
for testing the validity of the various theoretical models. 

In this contribution we present such a comparison in a 
systematic way. We consider for reference the 
\texttt{NNPDFpol1.1}~\cite{Nocera:2014gqa} 
and the \texttt{DSSV08}~\cite{deFlorian:2009vb} parton sets.
They both provide polarized PDFs at next-to-leading-order (NLO)
accuracy, but they differ in two main aspects:
first, the methodology used for PDF determination, and, second,
the experimental information included for constraining
$\Delta\bar{u}$ and $\Delta\bar{d}$ distributions.

Concerning the methodology used for PDF determination,
the \texttt{DSSV08} analysis is based on the \textit{standard}
Hessian methodology. Conversely, the \texttt{NNPDFpol1.1} analysis 
is based on the NNPDF methodology. 
This uses a parametrization of PDFs based 
on neural networks with a very large number of free parameters,
and Monte Carlo sampling and representation 
of PDFs. The first feature allows
for reducing as much as possible the theoretical bias due to PDF
parametrization. The second feature allows in 
particular for including new experimental information 
into a given PDF determination via Bayesian 
reweighting~\cite{Ball:2010gb,Ball:2011gg}. This
consists of updating the representation of the probability 
distribution in the space of PDFs, provided by an available PDF set,
by means of Bayes' theorem, in such a way that the 
information contained in the new data sets is included.
Indeed, the \texttt{NNPDFpol1.1} parton set was obtained via reweighting
(see Refs.~\cite{Nocera:2014gqa,Ball:2013lla} for details).

Concerning the experimental information included for constraining
$\Delta\bar{u}$ and $\Delta\bar{d}$ distributions,
the \texttt{NNPDFpol1.1} PDF determination uses 
$W$-boson production measurements in polarized \textit{pp} collisions. 
Specifically, the data sets on both $W^+$ and $W^-$ single- and double-spin 
asymmetries recently provided by the STAR 
collaboration~\cite{Adamczyk:2014xyw} were included.
Unlike \texttt{NNPDFpol1.1}, in the \texttt{DSSV08} analysis
sea-quark densities are determined from SIDIS data, 
since $W$-boson production data were not available at 
the time this global fit was performed. 
However, in this case additional knowledge 
of the fragmentation functions for light
quarks is required. Indeed, fragmentation functions 
are on the same footing as PDFs: 
they can only be determined from a fit to experimental data (see
\textit{e.g}~\cite{Albino:2008gy} and references therein), 
and as such they are subject 
to the same potential sources of bias. 
Because the NNPDF methodology aims at reducing
possible such bias, and a determination of fragmentation
functions based on this methodology is not yet available, 
in \texttt{NNPDFpol1.1} sea-quark distributions were determined 
from collider data only, which do not require the usage of fragmentation
functions. 

\begin{figure}[t]
\centering
\includegraphics[scale=0.35]{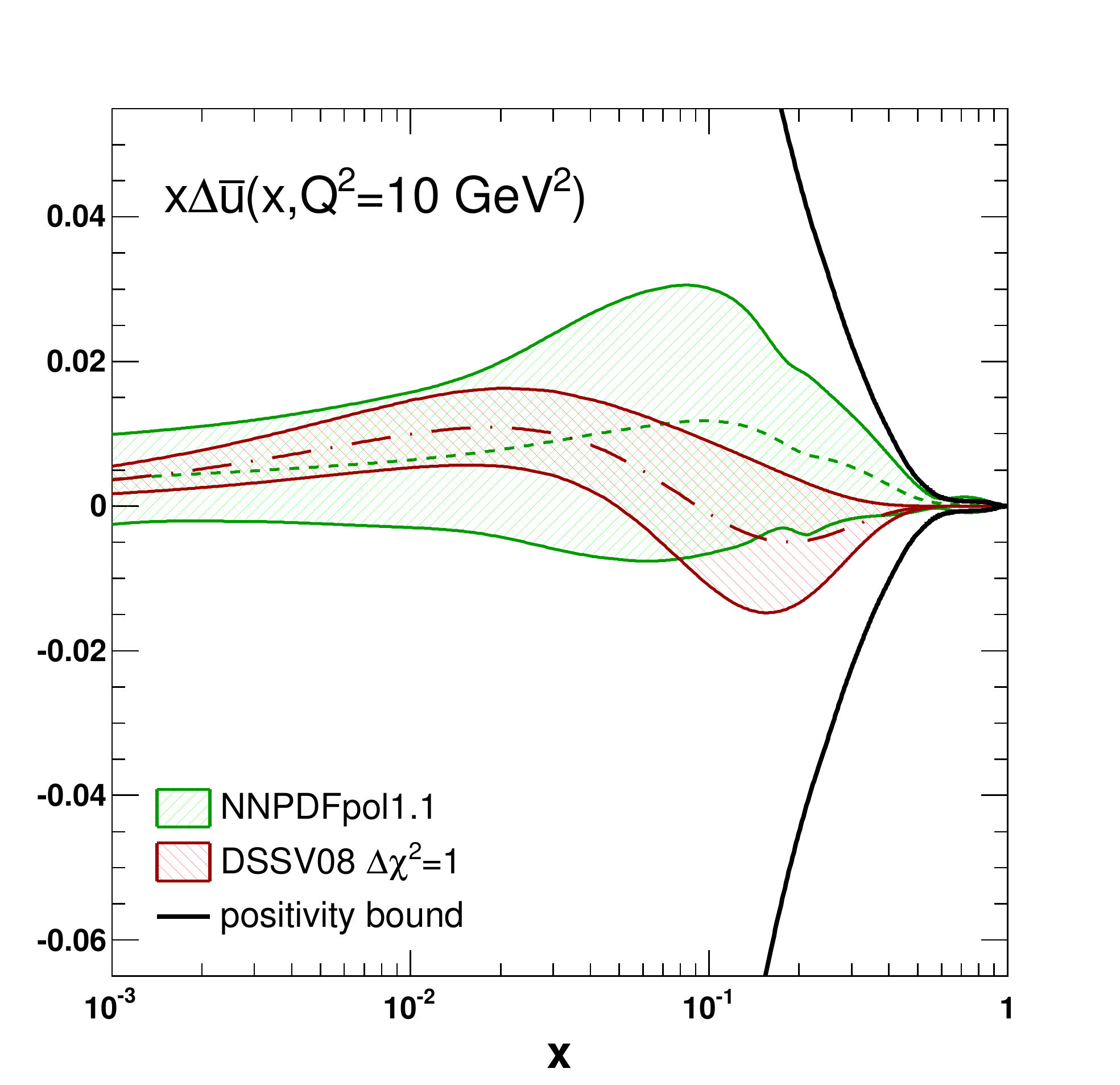}
\includegraphics[scale=0.35]{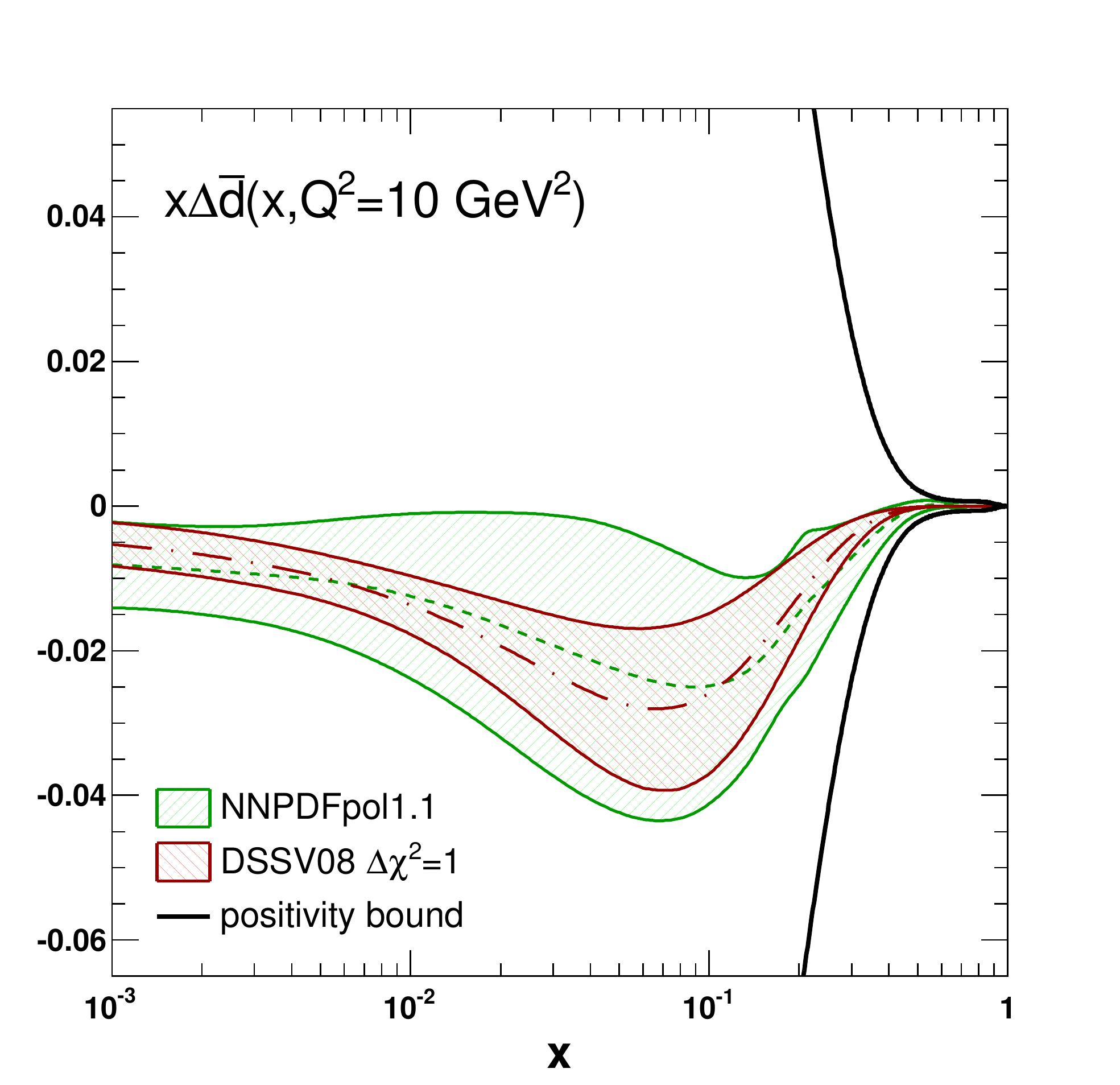}
\caption{The $\Delta\bar{u}$ and $\Delta\bar{d}$ distributions from the 
\texttt{NNPDFpol1.1}~\cite{Nocera:2014gqa} parton set compared to those
from the \texttt{DSSV08}~\cite{deFlorian:2009vb} parton set at $Q^2=10$~GeV$^2$. 
The positivity bound is also shown (see 
Refs.~\cite{Nocera:2014gqa,Ball:2013lla} for details).}
\label{fig:seaPDFs}
\end{figure}

Light sea-quarks from the \texttt{NNPDFpol1.1} parton set are shown in 
Fig.~\ref{fig:seaPDFs}, together with those from the \texttt{DSSV08} 
parton set; the \texttt{NNPDFpol1.1} uncertainty corresponds to the one-$\sigma$
band, while the \texttt{DSSV08} uncertainty is the Hessian
uncertainty computed assuming $\Delta\chi^2=1$. This choice may lead 
to somewhat underestimated uncertainties: it is well known that, in
global fits based on Hessian methodology, a tolerance $\Delta\chi^2=T>1$
is needed for faithful uncertainty estimation. Indeed, 
in Ref.~\cite{deFlorian:2009vb} uncertainty estimates obtained from 
the Lagrange multiplier method with $\Delta\chi^2/\chi^2=2\%$ 
(roughly corresponding to $T\sim 8$) were recommended as more
reliable. In this case, \texttt{DSSV08} uncertainties would be larger than those
shown in Fig.~\ref{fig:seaPDFs} by a factor $\sqrt{T}$.

The two results show a slightly different behavior of the 
$\Delta\bar{u}$ distribution above $x \sim3\cdot 10^{-1}$, 
while they nicely agree for the 
$\Delta\bar{d}$ distribution. Since light sea-quarks are obtained from 
different hard-scattering processes in the two determinations, 
$W$-boson production in \texttt{NNPDFpol1.1} and SIDIS in \texttt{DSSV08},
the discrepancy in the $\Delta\bar{u}$ distribution may suggest some tensions
between the corresponding experimental data. Also, the \texttt{DSSV08} result
could have been biased by an imperfect knowledge of the fragmentation 
functions used in the analysis of SIDIS data.

\begin{figure}[t]
\centering
\includegraphics[scale=0.35]{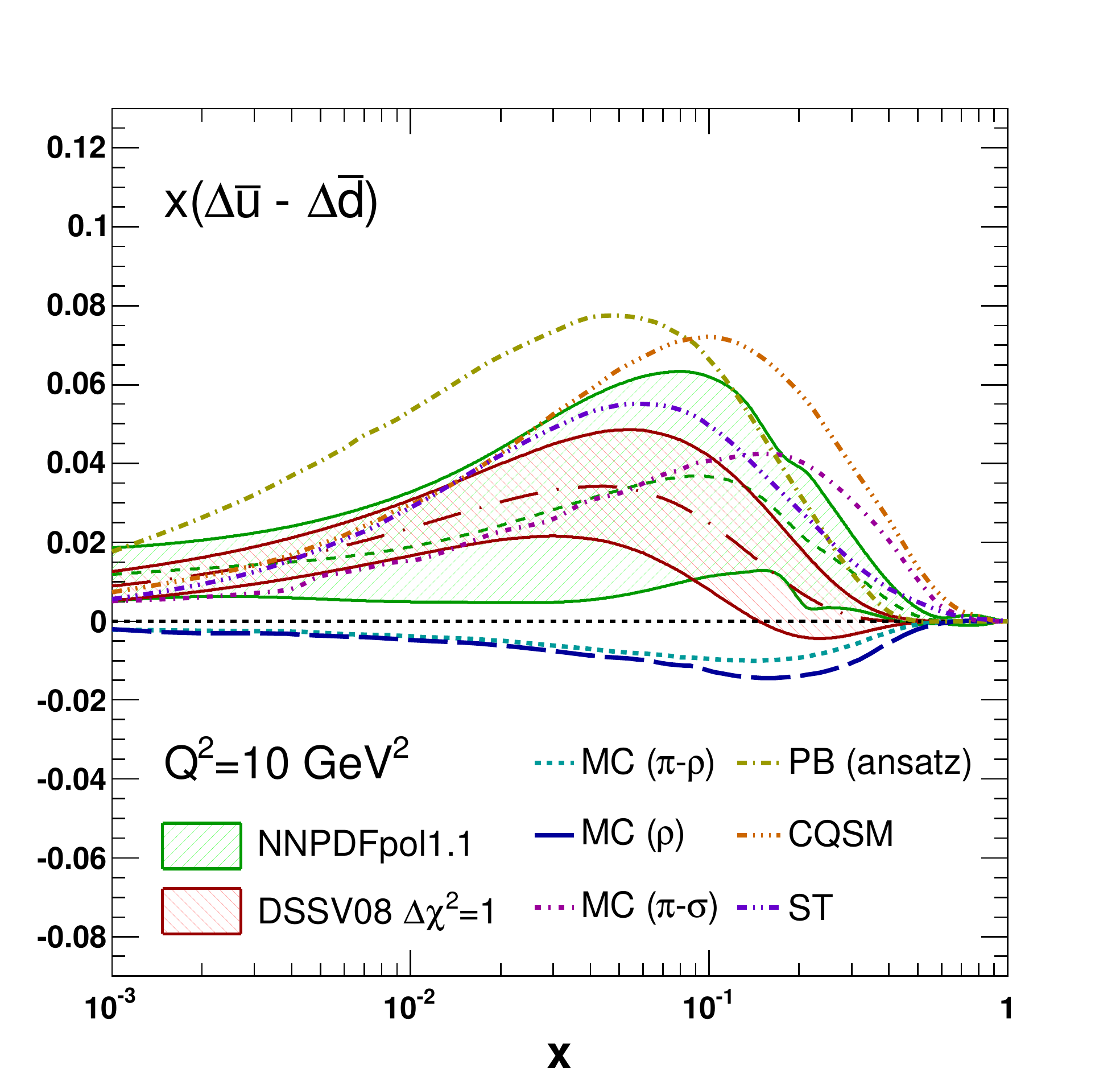}
\includegraphics[scale=0.35]{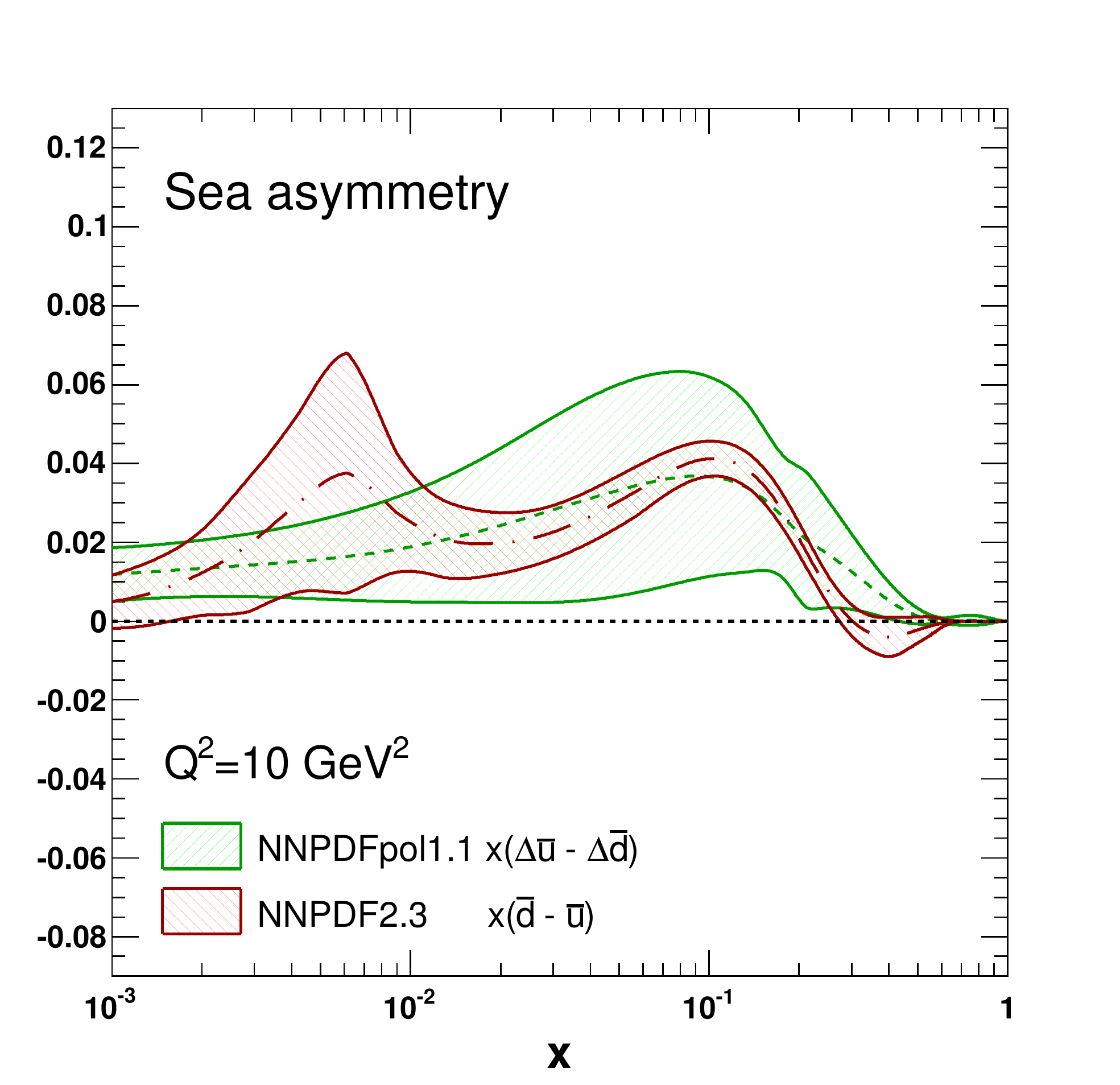}
\caption{The polarized light sea-quark asymmetry $x(\Delta\bar{u}-\Delta\bar{d})$
computed with the \texttt{NNPDFpol1.1} PDFs~\cite{Nocera:2014gqa} 
at $Q^2=10$~GeV$^2$, its comparison with predictions computed 
with \texttt{DSSV08}~\cite{deFlorian:2009vb} PDFs and various models of nucleon 
structure~\cite{Cao:2001nu,Kumano:2001cu,Fries:2002um,Gluck:2000dy,Dressler:1998zi,Wakamatsu:2014asa,Bhalerao:2001rn}
(left plot), and with the unpolarized asymmetry 
$x(\bar{d}-\bar{u})$ computed with \texttt{NNPDF2.3} PDFs~\cite{Ball:2012cx} (right plot).}
\label{fig:asysea}
\end{figure}

In Fig.~\ref{fig:asysea}, we show the polarized light sea-quark
asymmetry $x(\Delta\bar{u}-\Delta\bar{d})$ computed with the 
\texttt{NNPDFpol1.1} PDFs at $Q^2=10$~GeV$^2$. This result is 
compared with that obtained from the \texttt{DSSV08} PDFs and
from various models of nucleon structure, 
and with the unpolarized sea-quark asymmetry
$x(\bar{d}-\bar{u})$ computed with the PDFs from the 
\texttt{NNPDF2.3} parton set~\cite{Ball:2012cx}. 
In Tab.~\ref{tab:models}, we collect the values of the integrated
asymmetry $I_{\Delta}=\int_0^1dx[\Delta\bar{u}(x)-\Delta\bar{d}(x)]$
for various models, as reported in Ref.~\cite{Chang:2014jba}, 
compared to that obtained using the \texttt{NNPDFpol1.1} and \texttt{DSSV08} PDFs. 
The theoretical models quoted in Fig.~\ref{fig:asysea} and Tab.~\ref{tab:models}
include several meson-cloud (MC) 
models~\cite{Eichten:1991mt,Cao:2001nu,Kumano:2001cu,Fries:2002um},
various Pauli-blocking (PB) models~\cite{Cao:2001nu,Gluck:2000dy,Steffens:2002zn},
chiral quark-soliton (CQS) model~\cite{Dressler:1998zi,Wakamatsu:2014asa},
instanton (IN) model~\cite{Dorokhov:2001pz} and statistical (ST)
model~\cite{Bhalerao:2001rn}. A description of each of these models
has been recently summarized in Ref.~\cite{Chang:2014jba}.
\begin{table}[t]
 \centering
 \scriptsize
 \begin{tabular}{lcccccc}
  \toprule
               & MC ($\pi$-meson)~\cite{Eichten:1991mt} 
               & MC ($\rho$-meson)~\cite{Kumano:2001cu}
               & PB (bag-model)~\cite{Cao:2001nu}
               & PB~\cite{Steffens:2002zn}
               & IN~\cite{Dorokhov:2001pz}
               & \texttt{DSSV08} $\Delta\chi^2=1$~\cite{deFlorian:2009vb}\\
  $I_{\Delta}$ & $\equiv0$
               & $<0$
               & $\simeq 0.09$ 
               & $\simeq 0.2$
               & $\simeq 0.2$
               & $\mathbf{0.14\pm 0.05}$\\
  \midrule  
               & MC ($\pi$-$\rho$ inter.)~\cite{Cao:2001nu}
               & MC ($\pi$-$\sigma$ inter.)~\cite{Fries:2002um}
               & PB (ansatz)~\cite{Gluck:2000dy}
               & CQS~\cite{Dressler:1998zi,Wakamatsu:2014asa}
               & ST~\cite{Bhalerao:2001rn}
               & \texttt{NNPDFpol1.1}~\cite{Nocera:2014gqa}\\
  $I_{\Delta}$ & $[-4\cdot 10^{-3},-0.033]$ 
               & $\simeq 0.12$
               & $\simeq 0.3$
               & $0.31$
               & $>0.12$
               & $\mathbf{0.17\pm 0.08}$\\
  \bottomrule
 \end{tabular}
 \caption{Prediction for the quantity 
$I_{\Delta}=\int_0^1dx[\Delta\bar{u}(x)-\Delta\bar{d}(x)]$
based on various theoretical models. Results are
taken from Ref.~\cite{Chang:2014jba}. The value of $I_{\Delta}$ at 
$Q^2=10$ GeV$^2$, obtained using the \texttt{NNPDFpol1.1}~\cite{Nocera:2014gqa}  
and the \texttt{DSSV08}~\cite{deFlorian:2009vb} PDF determinations
from experimental data, is shown for comparison.}
 \label{tab:models}
\end{table}

Inspection of Fig.~\ref{fig:asysea} and Tab.~\ref{tab:models} allows
us to draw the following conclusions.
\begin{itemize}
 \item The results obtained using either the \texttt{NNPDFpol1.1}
 or the \texttt{DSSV08} PDFs are in fairly good agreement, for
 both the sea-quark asymmetry and its first moment $I_{\Delta}$.
 A slight difference in the size and shape of their central values 
 is observed, especially in the high-$x$ region, 
 as a consequence of the different behavior of the 
 $\Delta\bar{u}$  distribution in the two parton sets, 
 see Fig.~\ref{fig:seaPDFs}. This mild discrepancy 
 may be explained by the fact that,
 in comparison to \texttt{NNPDFpol1.1}, the \texttt{DSSV08} result 
 could be affected by some sources of bias: these include a poor 
 knowledge of the fragmentations functions used to analyze SIDIS data,
 and a parametrization of PDFs which is not sufficiently flexible.
 The uncertainty for both the asymmetry in Fig.~\ref{fig:asysea} 
 and its first moment $I_{\Delta}$ in Tab.~\ref{tab:models} corresponds
 to the one-$\sigma$ band (for \texttt{NNPDFpol1.1}) and to the Hessian 
 uncertainty computed assuming $\Delta\chi^2=1$ (for \texttt{DSSV08}).
 We notice that in Ref.~\cite{deFlorian:2009vb} a more faithful
 estimate for \texttt{DSSV08} uncertainties was provided, assuming 
 a tolerance $T=\Delta\chi^2\sim 8$ (see the discussion after Fig.~\ref{fig:seaPDFs}).
 In this case, \texttt{DSSV08} uncertainties are larger than those shown
 in both Fig.~\ref{fig:asysea} and Tab.~\ref{tab:models} by a factor $\sqrt{T}$.
 \item The polarized sea-quark asymmetry obtained from experimental data
 is definitely positive, within uncertainties, for both the 
 \texttt{NNPDFpol1.1} and \texttt{DSSV08} determinations. Hence, predictions
 from models of nucleon structure which lead to a negative asymmetry are
 clearly disfavored. This is the case of MC models of 
 Refs.~\cite{Eichten:1991mt,Cao:2001nu,Kumano:2001cu}. 
 While all other models are qualitatively consistent
 with both \texttt{NNPDFpol1.1} and \texttt{DSSV08} results,
 predictions from the MC and ST models of Refs.~\cite{Fries:2002um,Bhalerao:2001rn} 
 are in better agreement with the \texttt{NNPDFpol1.1} result than those 
 from the CQS and PB models of Refs.~\cite{Dressler:1998zi,Wakamatsu:2014asa,Gluck:2000dy}.
 A more precise experimental knowledge of the sea-quark asymmetry
 is needed to reduce its uncertainty and eventually discriminate between these models. 
 \item The polarized sea-quark asymmetry has approximately the same 
 absolute size as its unpolarized counterpart, but it is affected by a larger uncertainty 
 (see right plot in Fig.~\ref{fig:asysea}).
 Even within this uncertainty, it is clear that polarized and unpolarized asymmetries
 have opposite sign. However, it is difficult to decide whether the expectation 
 $|\Delta\bar{u} - \Delta\bar{d}|>|\bar{u} - \bar{d}|$, predicted for instance
 by the CQS model, is actually fulfilled. Note that this implies that the 
 Bjorken sum rule (which is proportional to the first moment of the triplet PDF 
 combination $\Delta T_3=\Delta u + \Delta\bar{u} - \Delta d - \Delta\bar{d}$) 
 is not entirely given by valence quarks: 
 also sea quaks can contribute to it, through their flavor asymmetry.
\end{itemize}

In conclusion we have shown, for the first time, that $W$-boson production data
favor $\Delta\bar{u}>0>\Delta\bar{d}$ and a positive flavor asymmetry
of polarized light antiquarks in the nucleon. This agrees with the result obtained using
SIDIS, and allows us to conclude that some theoretical models of nucleon structure
are disfavored. 
Ongoing experimental measurements on $W$ asymmetries at the Relativistic Heavy Ion Collider
will help to reduce the uncertainties on the polarized antiquarks:
more stringent constraints on the validity of theoretical models, and
a more clear picture of the flavor asymmetry of light antiquarks in the
nucleon could be then achieved.

\end{document}